\documentclass{DISproc}

\begin{document}
\title{
Measurement of Normalised Multi-jet Cross Sections using Regularised Unfolding and Extraction of $\mathbf{\alpha_s(M_Z)}$ in DIS at High $\mathbf{Q^2}$ }

\author{{\slshape Daniel Britzger$^1$}\\ on behalf of the H1 Collaboration\\[1ex]
$^1$DESY, Notkestra{\ss}e 85, 22607 Hamburg, Germany}

\contribID{xy}

\doi  

\maketitle

\begin{abstract}
New results on normalised inclusive jet, dijet and trijet differential cross sections in neutral current DIS based on a regularised unfolding procedure are presented.
Compared to a previously published result on normalised multi-jet cross sections, the new features are an extended range in jet pseudorapidity, an improved hadronic energy scale uncertainty of $1\%$ and the application of an unfolding procedure.
The normalised jet cross sections are compared to QCD calculations at NLO.
The value of the strong coupling determined from the normalised inclusive jet, dijet and trijet measurements simultaneously is $\alpha_s(M_Z) = 0.1163~\pm 0.0008~(\textrm{exp.})~\pm 0.0011~(\textrm{sys.}) ^{+0.0044}_{-0.0035}~(\textrm{theo.)}~\pm0.0014~(\textrm{PDF})$.
\end{abstract}

\section{Introduction}
Jet production in neutral current DIS at HERA provides an important testing ground for QCD.
The measurement of jet quantities is directly sensitive to the strong coupling $\alpha_s$ and can give constraints on the gluon density in the proton.
Furthermore, it is a valuable benchmark process for Monte Carlo event generators, particularly with regard to parton showers.
Two different kinds of jet measurements can be distinguished.
Inclusive jet measurements, where each single jet is counted, and jet measurements like dijet and trijet measurements, where each event that fulfills topological and kinematic criteria on jet quantities contributes to the cross section once.
Both approaches allow to extract the strong coupling by comparing to perturbative QCD predictions.\\

The measurement presented here is based on data with an integrated luminosity of $\unit{361}{\picobarn^{-1}}$ collected in the years 2003~-~2007 with the H1 detector~\cite{H1Prel-12-031}.
The data are identical to a previous H1 analysis of absolute jet cross sections~\cite{H1Prel-11-032}.
This analysis is extended to normalised cross sections, where the normalisation is performed with respect to the NC DIS cross section.
In~\cite{H1Prel-11-032} and in the analysis reported here, improvements on the reconstruction of tracks and the calorimetric energy were applied \cite{Kogler:2011}.
The correction of detector effects to determine the particle level cross section is performed using a regularised unfolding procedure and is presented in more detail in this document.


\section{Phase space definition}


The NC~DIS events are selected by requiring an identified scattered electron, a virtuality of the exchanged boson ($\gamma$/Z$^0$) of $150<Q^2<\unit{15000}{\GeV^2}$ and an inelasticity of the interaction of $0.2<y<0.7$.
The jet finding is performed in the Breit frame of reference where the exchanged boson is completely space-like.
Particle candidates of the hadronic final state are clustered into jets using the inclusive $k_\mathrm{T}$ algorithm \cite{Ellis:93:3160} with a distance parameter $R_0 = 1$, as implemented in the FastJet package \cite{Cacciari:2011ma}.
The jets are required to be in the pseudorapidity range in the laboratory rest frame between $-1.0<\mathrm{\eta^{jet}_{lab}}<2.5$, and the jet momentum in the Breit frame is required to be $7<P_{\mathrm{T}}<\unit{50}{\GeV}$.
Events with at least two (three) jets with transverse momentum larger than \unit{5}{\GeV} are considered as dijet (trijet) events if the two leading jets of the measured observables have an invariant mass $M_{jj}$ exceeding \unit{16}{\GeV}.
For events that fulfill the dijet (trijet) criteria, the average transverse momentum is $\langle P_{\mathrm{T}} \rangle = \frac{1}{N} \sum_i^N P_{\mathrm{T}}^{\mathrm{jet},i}$, with $N=2(3)$.\\
All three double differential jet measurements are normalised to the inclusive NC~DIS measurement as function of $Q^2$.
The advantage of the normalised jet cross sections compared to absolute jet cross sections~\cite{H1Prel-11-032} are reduced systematic uncertainties on the experimental as well as on the theoretical side.

\section{Detector correction using regularised unfolding}
Due to kinematic migrations because of resolution and other effects and due to the limited acceptance of the detector, the data have to be corrected.
For this purpose a multidimensional regularised unfolding procedure, including all correlations, is applied. This procedure makes use of a migration matrix $A$ that connects the \emph{particle level}, represented by the vector $x$, with the \emph{detector level}, represented by the vector $y$, such that the equation $y=Ax$ holds.
The particle level distribution $x$ is determined using the \textsc{TUnfold} package \cite{Schmitt:2012kp}. The $\chi^2$-function
\begin{equation}\label{eq:chi2unfolding}
\chi^2(x) = ( y - A \cdot x ) ^T~V^{-1}_{yy}~ ( y - A \cdot x ) + \tau^2(x-x_0)^T(L^TL)(x-x_0)
\end{equation}
is minimised analytically as a function of $x$, where $V_{yy}$ is the covariance matrix, and $x_0$ is the bias distribution.
The Tikhonov regularisation parameter $\tau$ protects the result from large fluctuations, and a regularisation condition for the matrix $L$ the unit matrix is chosen.\\

The unfolding is performed using a single matrix $A$ with an overall $4 \times 4$ structure that allows to unfold four measurements (NC DIS, inclusive jets, dijet, trijet) all at once and further gives the possibility to determine the normalised cross sections taking all correlations into account (see sec. \ref{Sec:Result}).
The four diagonal elements are submatrices that describe the migrations of just one single measurement.\\
\textbf{NC DIS submatrix:}
The migrations of the NC DIS measurements are described in two dimensions, i.e.\ in the kinematic variables $y$ and $Q^2$. The NC DIS data are used to determine the normalised jet cross sections.\\
\textbf{Inclusive jets submatrix:}
The inclusive jet measurement is a measurement of jet multiplicities, where jets are defined by the jet algorithm independently on particle level and on detector level. This implies that jets have to be connected between both levels.
A geometric jet matching method based on a closest pair algorithm with a distant measure $R = \sqrt{ \Delta\phi^2 + \Delta\eta^2 }$ and a maximum distance of $R<0.9$ is applied.
This ensures that no kinematical biases are introduced.
\begin{wrapfigure}{r}{0.60\textwidth}
  \centering
  \includegraphics[width=0.60\textwidth]{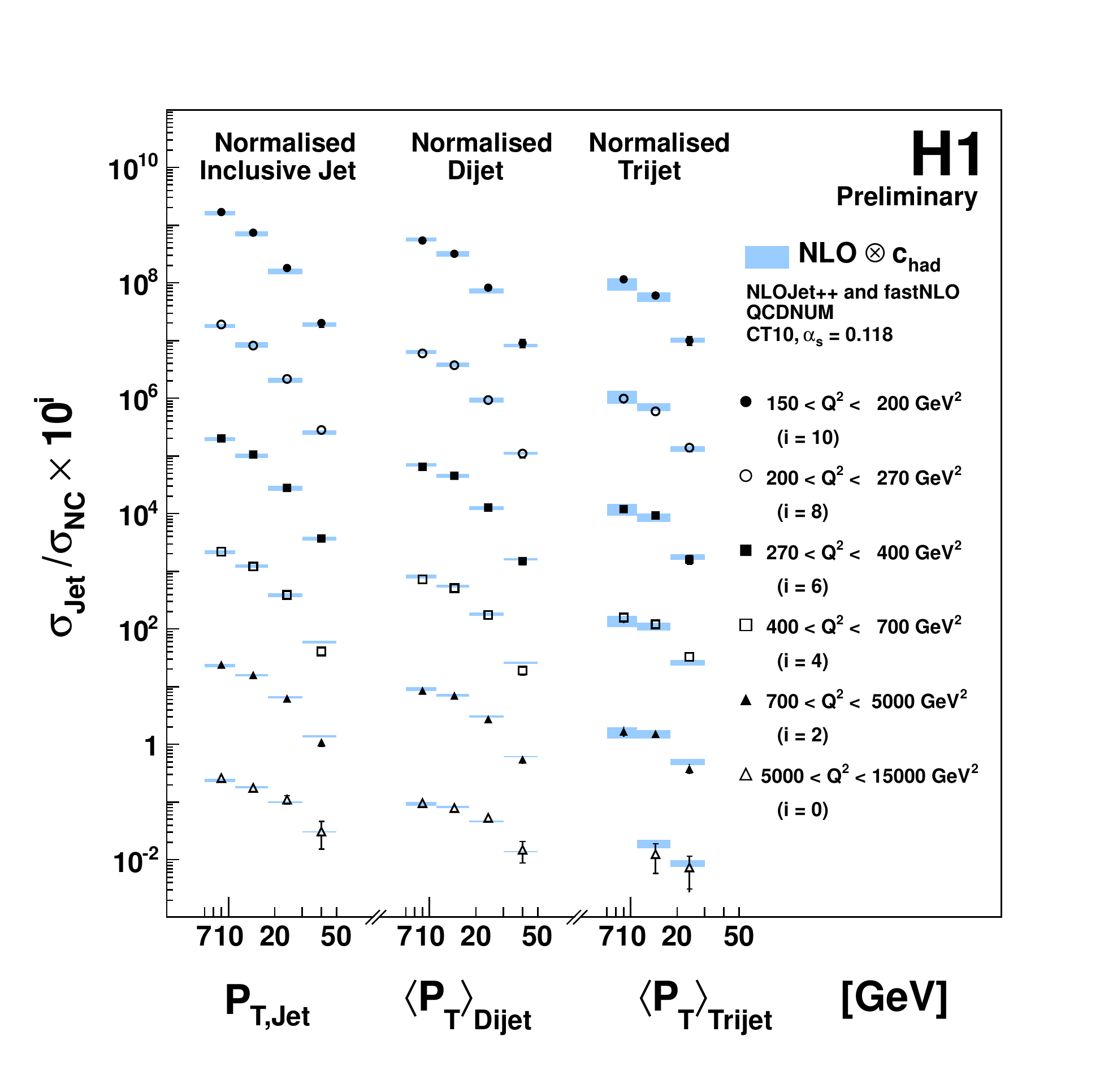}
  \caption{
  Normalised inclusive jet, dijet and trijet measurements in six bins of the virtuality of the exchanged boson $Q^2$ and in bins of the transverse jet momentum $P_{\mathrm{T}}$ in the Breit frame or the average transverse momentum of the two (three) leading jets $\langle P_{\mathrm{T}} \rangle$, respectively. The correlated statistical errors are shown by the inner error bars which are mostly smaller than the marker size.
  The outer error bar shows the total experimental uncertainty, incorporating all systematic experimental uncertainties.
  The data are corrected for detector effects using a regularised unfolding technique.
  The measurements are compared to next-to-leading order QCD calculations using the CT10 PDF set.
  They are corrected for hadronisation effects.
  The theory predictions include uncertainties determined by scale variations of a factor $2$ up and down.
   }
  \label{Fig:CrossSecions}
\end{wrapfigure}
Matched jets are filled into the submatrix of the inclusive jets in a three-dimensional unfolding scheme in the variables $P_{\mathrm{T}}$, $Q^2$ and $y$.
Jets that appear on generator level only are treated as inefficiencies.
Jets that appear on detector level only are difficult to handle and cause a large error on the overall normalisation if not handled properly.
Detector-level-only jets do not have particle level jet quantities, but still event observables on particle level are known.
These jets are filled into the submatrix that connects the detector level inclusive jets to the particle level NC~DIS measurement.
The normalisation of the NC DIS measurement is preserved by adding negative weights to the efficiency accordingly.\\
\textbf{Dijet and trijet submatrices:}
The unfolding of the dijet and trijet measurement is performed in a three dimensional unfolding scheme in the kinematic variables $\langle P_{\mathrm{T}} \rangle$, $Q^2$ and $y$.
Additional bins describe migrations in and out of the dijet and trijet phase space.
Migrations in the kinematic variables $M_{jj} > \unit{16}{\GeV}$, $P_{\mathrm{T}}^{\textrm{jet2}}>\unit{5}{\GeV}$ and requirements on $\mathrm{\eta_{lab}^{jet}}$ are taken into account.
Similarly as for the inclusive jets, events that do not fulfill the dijet (trijet) requirements on particle level but on detector level are estimated by the NC DIS events.\\

The covariance matrix on detector level $V_{yy}$ is determined by data and contains the statistical uncertainty and all correlations between the four measurements.
The unfolded covariance matrix $V_{xx}$ on particle level is determined by error propagation through the unfolding process and holds information on the correlations resulting from detector effects as well as the propagated correlations between the single measurements.

\section{Result}\label{Sec:Result}

\subsection{Normalised multijet cross sections}
Each bin of the jet measurements is normalised with the corresponding bin of the inclusive NC~DIS measurement.
The covariance matrix of all three normalised jet measurements is determined by a
full error propagation using the covariance matrix $V_{xx}$.
Also the systematic errors are determined by a full error propagation.
This procedure is equivalent to a direct measure of normalised jet cross sections.
The normalised inclusive jet, dijet and trijet measurements are shown in Fig. \ref{Fig:CrossSecions}, where they are compared to pQCD predictions which show a good agreement over the full phase space.

\subsection{Determination of the strong coupling constant}
The strong coupling is determined by performing a $\chi^2$-minimisation procedure to all three normalised jet measurements simultaneously with $\alpha_s(M_Z)$ as a free parameter.
The theory calculations are performed using the \textsc{QCDNUM} program \cite{Botje:2010ay} for the NC~DIS cross sections and the \textsc{NLOJet++} program \cite{Nagy:99,Nagy:01} interfaced to \textsc{FastNLO} \cite{Wobisch:2011ij,Britzger:12} for a fast repeated calculation of the jet cross sections using the CT10 PDF set \cite{Lai:2010vv}.
The $\chi^2$-definition includes the full covariance matrix after the unfolding and takes into account systematic uncertainties using nuisance parameters.
In order to consider only bins that show a fast convergence, bins with large $k$-factors of $k>1.3$ are excluded from the fit, where $k = \sigma_{\textrm{NLO}}/\sigma_{\textrm{LO}}$.
The resulting fit takes contributions from $42$ out of $65$ bins into account and shows a reasonable $\chi^2 / \textrm{ndf}$ of $53/41$. The resulting $\alpha_s({M_Z})$ is determined to be
\begin{displaymath}
    \alpha_s(M_Z) = 0.1163 \pm 0.0008 ~\mathrm{(exp.)} \pm 0.0011  ~\mathrm{(had.)} \pm 0.0014 ~\mathrm{(pdf)}^{~+0.0044}_{~-0.0035}~\mathrm{(theo.)}.
\end{displaymath}
This result is consistent with previous H1 publications of normalised multi-jet cross sections \cite{Aaron:2009vs}.
The uncertainties from hadronisation corrections (had.), the PDF uncertainty (pdf) and theoretical uncertainties from missing higher orders (theo.) are determined by repeating the fit with shifted theory cross sections.
Each of the theoretical uncertainties is larger than the total experimental uncertainty.
The result is consistent with the world average within the errors~\cite{Bethke:09:689}.




{\raggedright
\begin{footnotesize}


\bibliographystyle{DISproc}
\bibliography{britzger_daniel_h1jets}

\providecommand{\href}[2]{#2}\begingroup\raggedright\begin{thebibliography}{10}

\bibitem{H1Prel-12-031}
{H1 Collaboration}.
\newblock preliminary result: H1prelim-12-031 (2012) .

\bibitem{H1Prel-11-032}
{H1 Collaboration}.
\newblock preliminary result: H1prelim-11-032 (2011) .

\bibitem{Kogler:2011}
R.~Kogler.
\newblock PhD thesis, Universit{\"a}t Hamburg, DESY-THESIS-2011-003,
  MPP-2010-175, 2011.

\bibitem{Ellis:93:3160}
S.~D. Ellis and D.~E. Soper.
\newblock Phys. Rev. D {\bfseries 48} (1993) 3160.

\bibitem{Cacciari:2011ma}
M.~Cacciari, G.~P. Salam, and G.~Soyez.
\newblock Eur.Phys.J. {\bfseries C72} (2012) 1896,
\href{http://arxiv.org/abs/1111.6097}{{\ttfamily arXiv:1111.6097 [hep-ph]}}.

\bibitem{Schmitt:2012kp}
S.~Schmitt.
\newblock
\href{http://arxiv.org/abs/1205.6201}{{\ttfamily arXiv:1205.6201
  [physics.data-an]}}.

\bibitem{Botje:2010ay}
M.~Botje.
\newblock
  \href{http://dx.doi.org/10.1016/j.cpc.2010.10.020}{Comput.Phys.Commun.
  {\bfseries 182} (2011) },
\href{http://arxiv.org/abs/1005.1481}{{\ttfamily arXiv:1005.1481 [hep-ph]}}.

\bibitem{Nagy:99}
Z.~Nagy and Z.~Trocsanyi.
\newblock Phys. Rev. D {\bfseries 59} (1999) 14020.

\bibitem{Nagy:01}
Z.~Nagy and Z.~Trocsanyi.
\newblock Phys. Rev. Lett. {\bfseries 87} (2001) 82001.

\bibitem{Wobisch:2011ij}
{M.~Wobisch~{\it et al.}}
\newblock
\href{http://arxiv.org/abs/1109.1310}{{\ttfamily arXiv:1109.1310 [hep-ph]}}.

\bibitem{Britzger:12}
{D. Britzger {\it et al.}}
\newblock these proceedings (2012) .

\bibitem{Lai:2010vv}
{H.-L.~Lai {\it et al.}}
\newblock \href{http://dx.doi.org/10.1103/PhysRevD.82.074024}{Phys.Rev.
  {\bfseries D82} (2010) 074024},
\href{http://arxiv.org/abs/1007.2241}{{\ttfamily arXiv:1007.2241 [hep-ph]}}.

\bibitem{Aaron:2009vs}
F.~Aaron {\em et~al.}
\newblock \href{http://dx.doi.org/10.1140/epjc/s10052-009-1208-7}{Eur.Phys.J.
  {\bfseries C65} (2010) },
\href{http://arxiv.org/abs/0904.3870}{{\ttfamily arXiv:0904.3870 [hep-ex]}}.

\bibitem{Bethke:09:689}
S.~Bethke.
\newblock Eur. Phys. J. C {\bfseries 64} (2009) 689.

\end{thebibliography}\endgroup
\end{footnotesize}
}


\end{document}